\documentclass[11pt]{article}
\usepackage{array, booktabs}
\usepackage{bm}
\usepackage{graphicx}
\usepackage{caption}
\usepackage{amsmath,amssymb,wasysym}
\usepackage{algorithm}
\usepackage{multirow}
\usepackage{algpseudocode}
\usepackage{setspace}  

\usepackage[utf8]{inputenc}
\usepackage{hyperref}
\usepackage[margin=1 in]{geometry}
\usepackage{subcaption}
\usepackage[normalem]{ulem}

\makeatletter
\newcommand*{\centerfloat}{%
  \parindent \z@
  \leftskip \z@ \@plus 1fil \@minus \textwidth
  \rightskip\leftskip
  \parfillskip \z@skip}
\makeatother

\usepackage{xcolor}

\usepackage[
    backend=biber,
    style=nature,
    defernumbers=true
]{biblatex}
\title{Noise-Induced Collective Memory \\
in Schooling Fish}
\author{Alyssa Chan$^{1,2}$ and Eva Kanso$^{1,2}$\footnote{Corresponding author: Kanso@usc.edu}}

\date{%
\footnotesize{
    $^1$Department of Aerospace and Mechanical Engineering,  \\ University of Southern California, Los Angeles, California 90089, USA\\%
    $^2$ Department of Physics and Astronomy,  \\ University of Southern California, Los Angeles, California 90089, USA\\%
    \today}
}

\usepackage{xr}
\makeatletter

\newcommand*{\addFileDependency}[1]{
\typeout{(#1)}
\@addtofilelist{#1}
\IfFileExists{#1}{}{\typeout{No file #1.}}
}\makeatother

\newcommand*{\myexternaldocument}[1]{%
\externaldocument{#1}%
\addFileDependency{#1.tex}%
\addFileDependency{#1.aux}%
}
 
\myexternaldocument{Supp}

\begin{document}

\maketitle
\begin{abstract}
Schooling fish often self-organize into a variety of collective patterns, from polarized schooling to rotational milling. Mathematical models support the emergence of these large-scale patterns from local decentralized interactions, in the absence of individual memory and group leadership.  In a popular model where individual fish interact locally following rules of avoidance, alignment, and attraction, the group exhibits collective memory: changes in individual behavior lead to emergent patterns that depend on the group's past configurations. However, the mechanisms driving this collective memory remain obscure.  Here, we combine numerical simulations with tools from bifurcation theory to uncover that
the transition from milling to schooling in this model is driven by a noisy transcritical bifurcation where the two collective states intersect and exchange stability.  We further show that key features of the group dynamics - the bifurcation character, transient milling, and collective memory - can be captured by a phenomenological model of the group polarization. Our findings demonstrate that collective memory arises from a noisy bifurcation rather than from structural bistability, thus resolving a long-standing ambiguity about its origins and contributing  fundamental understanding to collective phase transitions in a prevalent model of fish schooling.

\end{abstract}

\maketitle

\section{Introduction}

Animal groups such as fish schools and bird flocks transition fluidly between distinct collective patterns~\cite{Couzin2002, Cavagna2010}. This remarkable synchrony of the group motion emerges without a leader, as individuals make decisions based on locally acquired cues about the motion of others in the group~\cite{Couzin2007,Couzin2009,Heins2024}. Although each individual has spatially-limited sensing abilities and no awareness of the informational state of others, such as whether they know about the presence of a resource or a threat, behavioral coupling among neighbors allows pertinent information to propagate within the group, resulting in an effective range of perception much larger than the individual's actual sensory range~\cite{Couzin2007,Couzin2009,Cavagna2010,Cavagna2018}. These functional characteristics are shared across a wide range of animal group types, from insects~\cite{Buhl2006, Attanasi2014collective, Hensley2023,Sayin2025,Georgiou2025} to birds~\cite{Ballerini2008, Cavagna2010, Attanasi2014,Papadopoulou2022}, and even humans in a crowd~\cite{Moussaid2011, Moussaid2012, Xu2025, Rio2014,Gu2025}, 
and they have been reliably reproduced in a wide array of mathematical models that bridge the scale from individual to group behavior~\cite{Reynolds1987, Vicsek1995, Couzin2002,Calovi2014,Filella2018, Huang2024, Castro2024,Hang2025}.

\begin{figure}[!t]
    \centering
    \includegraphics[width=1\textwidth]{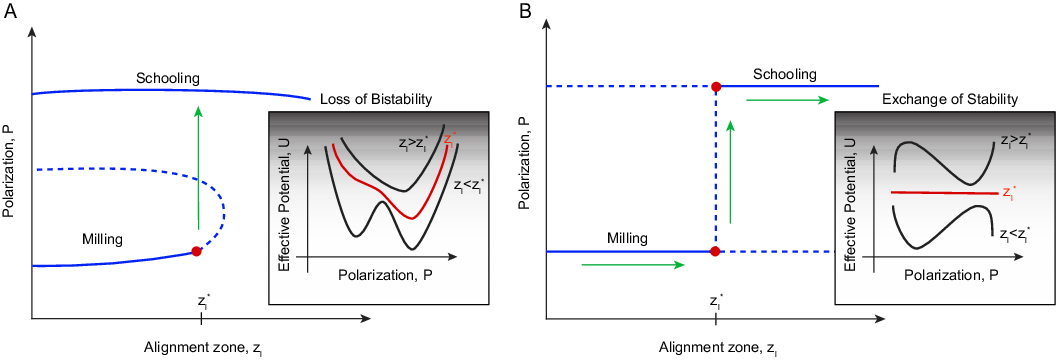}
    \caption{\footnotesize{Two possible mechanisms for the transition between milling and schooling as represented by the dependence of group-level polarization $P$ on individual-level parameter $z_l$: \textbf{A.} Subcritical pitchfork bifurcation: $P$ admits two equilibria, $P = 1$, which persists for all $z_l$ and $P = 0$, which exist only for $z_l$ below the bifurcation point $z_l^\ast$.  As a result, the system is bistable for $z_l<z_l^\ast$ and becomes monostable at $z_l>z_l^\ast$, where only $P=1$ is stable. This form of bistability implies that the dynamics is governed by a nonlinear double-well energy potential, which transitions to a single-well potential at $z_l^\ast$ and remains a single-well for $z_l>z_l^\ast$.
    \textbf{B.} Transcritical bifurcation: $P$ admits two equilibria at $0$ and $1$ that persist for all $z_l$ values, but intersect and exchange stability at the bifurcation point $z_l^\ast$, such that at $z_l<z_l^\ast$, $P=0$ is stable and $P=1$ is unstable, and vice versa for $z_l>z_l^\ast$. This bifurcation only maintains a single-well energy potential, whose minimum is governed by the individual parameter $z_l$.  For $z_l<z_l^\ast$, potential exist at one minimum, which flattens at $z_l^\ast$, and then shifts to a new minimum for $z_l>z_l^\ast$.
    }}
    \label{fig:map}
\end{figure}

In fish schools, collective patterns range from disordered swarming to more ordered phases such as rotational milling and polarized schooling~\cite{Radakov1973,Partridge1982}. These collective phases emerge in self-propelled particle models constructed using simple response rules, such as the popular “three-A rules” of avoidance, alignment, and attraction~\cite{Couzin2002}. Later models improved these strategies by inferring behavioral rules from tracking the trajectories of fish in a tank~\cite{Gautrais2012, Calovi2014}, incorporating flow interactions among the fish~\cite{Filella2018}, and accounting for interactions with domain boundaries~\cite{Gautrais2012,Huang2024}. Considerable attention was devoted to constructing phase spaces that map the individual behavior (parameters at the particle level) to the emergent collective phases~\cite{Calovi2014,Filella2018, Wang2022,Huang2024}. 

In establishing these functional maps from individual to collective behavior, bistability -- co-existence of two collective phases for the same individual behavior -- is crucial for understanding the dynamical properties of the group.  There is empirical support for collective bistability in groups of fish interacting with domain boundaries~\cite{Tunstrom2013,Lafoux2024} and under varying light intensities~\cite{Lafoux2023}.  In the empirically-derived self-propelled particle models~\cite{Gautrais2012}, bistability arises  under at least one of the following two conditions: when individual memory is incorporated in the model in the form of rotational inertia of the individual~\cite{Calovi2014,Wang2022} and when individuals are spatially confined~\cite{Kolpas2007,Huang2024}. In the latter, collective bistability is structural. 

Structural bistability arises when the group dynamics is governed by a double-well effective energy potential as demonstrated in Fig.~\ref{fig:map}A. Loss of bistability is associated with a bifurcation that causes one of the local minima to either change its stability, say via a pitchfork bifurcation, or to vanish altogether in a saddle-node bifurcation~\cite{Kolpas2007, Radisson2023,Huang2024}. 
Mechanistic insights into the energy landscape and bifurcations underlying the collective phase transitions provide a unifying framework for understanding the decision-making processes that are shared by animal groups and other complex systems, such as neural networks. Indeed, in both individual neurons and neural networks, the co-existence of multiple collective states at the same value of the system’s parameters (of the form illustrated in Fig.~\ref{fig:map}A) serve as key mechanisms for memory storage and temporal pattern recognition~\cite{Rabinovich2006, Sompolinsky1988,Lee2023,Mastrogiuseppe2018,Miller2010,Boaretto2021,Vecoven2021}. 


In the context of the three-A model, bistability depends on past states of the group~\cite{Couzin2002,Couzin2009}: slowly varying the individual behavior from a milling phase tends to maintain milling, while reversing the individual behavior back from a schooling phase tends to sustain polarized schooling. This form of bistability, termed hysteresis, is associated with collective memory and likened to multistability in neural systems~\cite{Couzin2007,Couzin2009,Sridhar2021}. However, the mechanisms giving rise to this hysteretic behavior are not known. Although hysteresis is commonly associated with structural bistability, it may also emerge from alternative mechanisms that do not involve multiple stable states. Take for example the dynamical system shown in Fig.~\ref{fig:map}B: two equilibria intersect and exchange stability at a critical parameter value. Known as a transcritical bifurcation, this mechanism could induce hysteresis in the presence of stochastic fluctuations, without ever exhibiting structural bistability~\cite{Arnold1998,Strogatz2018,Amir2023}. 



Here, we ask a fundamental, yet unexplored, question: is the collective memory, or hysteresis, reported in the three-A model truly a result of structural bistability of the form shown in Fig.~\ref{fig:map}A, thereby justifying its analogy to memory processes in the nervous system?  To address this question, we combine numerical simulations of the three-A model with analytical tools from bifurcation theory and stochastic differential equations. We find no evidence supporting a pitchfork bifurcation as in
Fig.~\ref{fig:map}A. Our analysis demonstrates that hysteresis in the three-A model arises from stochastic fluctuations near a transcritical bifurcation of the form illustrated in Fig.~\ref{fig:map}B. 


\section{Mathematical Model}

\paragraph{Vision-based Behavioral Rules.}
We consider a group of $N$ fish, where each individual is represented as a self-propelled particle moving at a constant speed $U$ in a three-dimensional space and following vision-based rules of \textit{avoidance}, \textit{alignment}, and \textit{attraction} to nearby neighbors \cite{Couzin2002}. Accordingly, an individual fish $i$ ($i = 1, \ldots N$) -- described by its position vector $\mathbf{r}_i(t)$ and heading direction $\mathbf{p}_i(t)$ as a function of time $t$ -- changes its heading $\mathbf{p}_i(t)$ based on the average position and direction of its neighbors.  The top priority of each fish is to reorient away from neighbors within its repulsion zone, or zone of \textit{avoidance}, defined as a spherical domain of radius $z_r$ centered at the focal fish. This ensures a minimum distance $z_r$ between individuals, approximately corresponding to the body length of the individual fish.  If no neighbors are present in the repulsion zone, the focal fish allocates equal attention to neighbors in its \textit{zone of alignment} and  \textit{zone of attraction}, defined as the spherical shells between the concentric spheres of radii $z_r$ and $z_l$, and of radii $z_l$ and $z_a$, respectively. For concreteness, let $n_r$ denote the number of neighbors in the repulsion zone of the focal fish $i$ such that the distance $\| \mathbf{r}_j - \mathbf{r}_i\| \leq z_r$ for $j\neq i$, $n_o$ the number of neighbors in the alignment zone for which
$z_r < \| \mathbf{r}_j - \mathbf{r}_i\| \leq z_l$, and $n_a$ the number of neighbors in the attraction zone for which $z_l < \| \mathbf{r}_j - \mathbf{r}_i\| \leq z_a$. The focal fish updates its heading direction at time $t + \Delta t$, where $\Delta t$ is a small time increment, according to
\begin{equation}
\begin{split}
\left. \mathbf{p}_i(t+\Delta t)\right|_{\rm no \ noise} =
    \begin{cases}
    - \left[\frac{1}{n_r} \sum_{j \neq i}^{n_r} \hat{\mathbf{r}}_{ij}(t)\right]/\|\cdot\|, & \quad \text{if } n_r \neq 0 \\[4ex]
       \left[\dfrac{\frac{1}{n_o}\sum_{j \neq i }^{n_o}  \mathbf{p}_j(t)}{\| \frac{1}{n_o}\sum_{j \neq i }^{n_o}  \mathbf{p}_j(t)\|} +  \dfrac{\frac{1}{n_a}\sum_{j \neq i}^{n_a}\hat{\mathbf{r}}_{ij}(t)}{\|   \frac{1}{n_a}\sum_{j \neq i}^{n_a}\hat{\mathbf{r}}_{ij}(t)\|}\right]/\| \cdot \|, & \quad \text{if } n_r = 0.
    \end{cases}
    \label{eq:vision}
\end{split}
\end{equation}
Here, $\hat{\mathbf{r}}_{ij} = (\mathbf{r}_j - \mathbf{r}_i) / \|\mathbf{r}_j - \mathbf{r}_i\|$ is the unit vector from individual $i$ in the direction of neighbor $j$, and $\| \cdot \|$ denotes the normalization of the vector quantity in the square brackets in the numerator.  We emphasize that individuals always prioritize the rule of repulsion before considering alignment and attraction. If there are no neighbors in either of the alignment or attraction zones, that is, if either $n_o$ or $n_a$ is equal to zero, the corresponding term in~\eqref{eq:vision} is discounted. If the resulting heading vector $\mathbf{p}_i(t+\Delta t)$ is zero, or if
there are no neighbors in all zones, the individual retains its current heading direction vector $\mathbf{p}_i(t+\Delta t) = \mathbf{p}_i(t)$.

Importantly, we account for the fact that live animals have a limited field of vision and, thus, we restrict each individual to see neighbors within an angular domain of size $\alpha$, characterized by a blind cone of angle $(360^\circ - \alpha)$ behind the individual. This limits the ability of the individual to perceive all neighbors in the alignment and attraction zones: neighbors within the blind zone are not detected and thus not accounted for in the last two sums in~\eqref{eq:vision}.  

\paragraph{Rotational Noise.} In addition to the individual's tendency to avoid, attract, and align with its neighbors, we add stochastic effects to the heading direction of each fish. Stochasticity here models the tendency of the individual to deviate from the vision-based rules in~\eqref{eq:vision} dictated by the positions and headings of its neighbors, whether deliberately as a result of the individual's “free will”~\cite{Filella2018} or due to sensory and response noise~\cite{Jiao2023}. 

\begin{table}[!b]
\centering
\caption{Model Parameters}
\begin{tabular}{l >{$}l<{$} l}
\toprule
\textbf{Parameter} & \textbf{Symbol} & \textbf{Values}\\
\midrule
Number of individuals & N & 100 \\
Radius of avoidance zone  & z_r & 1 body length\\
Radius of alignment zone & z_l & 1-7 body lengths\\
Radius of attraction zone & z_a & 4-14 body lengths\\
Field of vision & \alpha & 270$^o$ \\
Maximum turning angle & \theta_{\rm max} & 40$^o$ \\
Swimming speed & U & 3 body lengths per second\\
Noise intensity & \sigma & 0.01-0.10 \\
\bottomrule
\end{tabular}
\label{tab:param}
\end{table}

Noise is implemented by sampling a new direction from a spherically wrapped Gaussian centered on $\left. \mathbf{p}_i(t+\Delta t)\right|_{\rm no \ noise}$. Specifically, to each component of the unit vector $\left. \mathbf{p}_i(t+\Delta t)\right|_{\rm no \ noise}$, we add a random component sampled from a Gaussian distribution with standard deviation $\sigma$, then we re-normalize the resulting vector to one. Here, the standard deviation of the angular deviation between $\left. \mathbf{p}_i(t+\Delta t)\right|_{\rm no \ noise} $ and $\mathbf{p}_i(t+\Delta t)$ is nearly equal to $\sigma\sqrt{2/3}$, with $\sigma$ being the standard deviation of the Cartesian noise in each direction.

\paragraph{Rotational Constraint.} Biomechanics prevents an individual fish from performing sudden large turns~\cite{Jiao2023,Dunn2016,Voesenek2019}.  We thus consider a maximum turning angle $\theta_{\rm max}$ that constrains the degree of rotation between the current and desired heading directions, such that, if the angle between $\mathbf{p}_i(t+\Delta t)$ and $\mathbf{p}_i(t)$ exceeds $\theta_{\rm max}$, the individual rotates by at most $\theta_{\rm max}$ in the desired direction.

\begin{figure}[!t]
    \centering
    \includegraphics[width=1\textwidth]{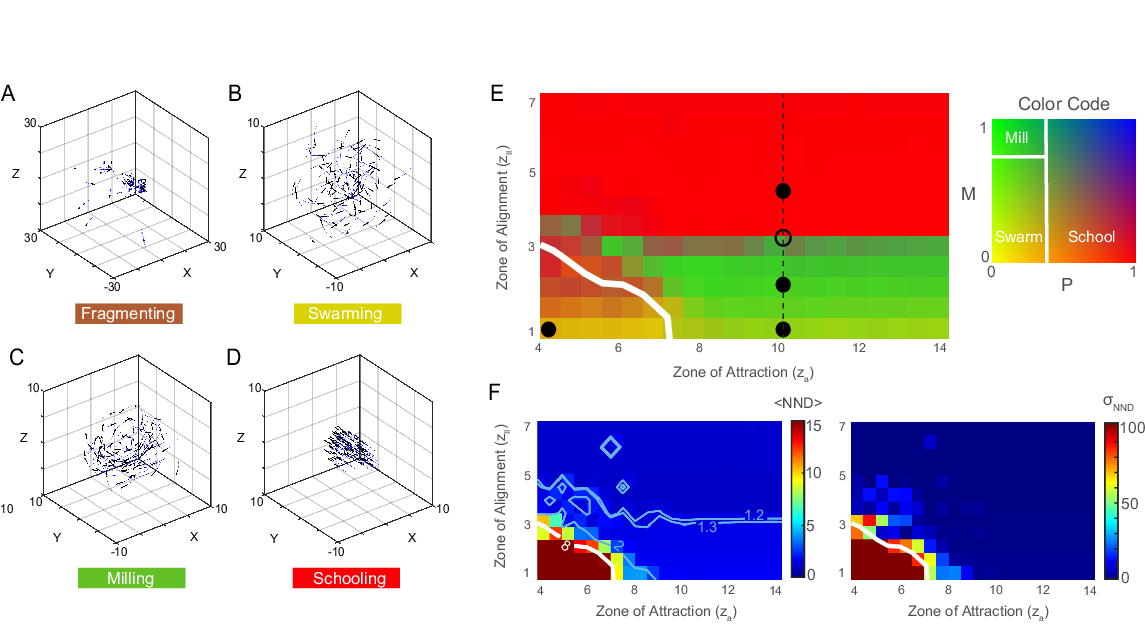}
    \caption{\footnotesize{
    Collective behavior of the fish group: \textbf{A.} Fragmentation at $z_l = 1.0$ and $z_a = 4.0$, \textbf{B.} Swarming at $z_l = 1.0$ and $z_a = 10.0$, \textbf{C.} Milling at $z_l = 2.0$ and $z_a = 10.0$, \textbf{D.} Schooling at $z_l = 4.0$ and $z_a = 10.0$.  
    \textbf{E.} Phase diagram as a function of alignment zone $z_l$ and attraction zone $z_a$, evaluated at steady-state using the group polarization $P$ and rotation $M$, identified with the two-dimensional color map to the right. Black markers mark the parameters used in A–D; open marker highlights the milling-to-schooling transition; dashed black line indicates $z_a = 10$ (see Fig~\ref{fig:bifurcation}).  \textbf{F.} Average nearest-neighbor distance (NND) and standard deviation $\sigma_{\rm NND}$ of NND; white line in E-F marks NND=$8$, beyond which the average NND exceeds the interaction range $z_a$ and exhibits large standard deviations, indicating school fragmentation.  Each value in E-F represents an average of 50 Monte Carlo (MC) simulations, with initial conditions randomly sampled from a three-dimensional unit cube. In all simulations, noise intensity $\sigma = 0.01$ and total integration time $T=500$.
    }}
    \label{fig:phase diagram}
\end{figure}

\paragraph{Statistical Order Parameters.}

To characterize the collective behavior of the group, we use two statistical order parameters: the group polar order parameter or \textit{polarization} $P$, and the group rotational or \textit{milling} order parameter $M$,   
\begin{align}
P(t) =  \left\| \frac{1}{N} \sum_{i=1}^N \mathbf{p}_i(t),   \right\|, \qquad  M(t) = \frac{1}{N} \left\| \sum_{i=1}^N \dfrac{(\mathbf{r}_{i}(t) -\mathbf{r}_{\rm c}(t))\times \mathbf{p}_i(t)}{\left\|(\mathbf{r}_{i}(t) -\mathbf{r}_{\rm c}(t))\times \mathbf{p}_i(t)\right\|} \right\|.
\end{align} 
Polarization $P$ quantifies the degree of alignment among individuals, with values ranging from $0$, indicating random movement with no alignment, to $1$, indicating perfect alignment with all individuals moving in the same direction.
Rotational order $M$ quantifies the collective angular momentum of all individuals about the group center $\mathbf{r}_{\rm c} = \frac{1}{N} \sum_{i=1}^N \mathbf{r}_i(t)$.
The values of $M$ ranges from $0$, indicating no net rotational order, to $1$, indicating maximal coordinated rotational motion about the group center.

\section{Results}

\paragraph{Emergent Collective Patterns.} We considered a school of 100 fish and varied the size of the alignment and attraction zones $z_l$ and $z_a$, while fixing the repulsion zone to $z_r = 1$; see~Table~\ref{tab:param} for the full list of parameters. Initially, the fish are randomly distributed in a three-dimensional cube of unit length. Depending on $z_l$ and $z_a$, the group of fish converges to one of four distinct collective patterns (Fig.~\ref{fig:phase diagram}A-D): at  $z_l = 1.0$ and $z_a=4.0$, the school fragments into subgroups; at $z_l = 1.0$ and $z_a=10.0$,  a disordered swarming phase emerges where the group remains cohesive but exhibits low polarization $P$ and rotational order $M$; at $z_l = 2.0$ and $z_a=10.0$, the group reaches a milling phase, forming a rotational “vortex” around the center of the group, with high rotational order $M$ and low polarization $P$; and at  $z_l = 4.0$ and $z_a=10.0$, all individuals swim in the same direction in a parallel schooling phase at high polarization $P$ and low rotational order $M$.

\begin{figure}[!t]
    \centering
    \includegraphics[scale = 1]{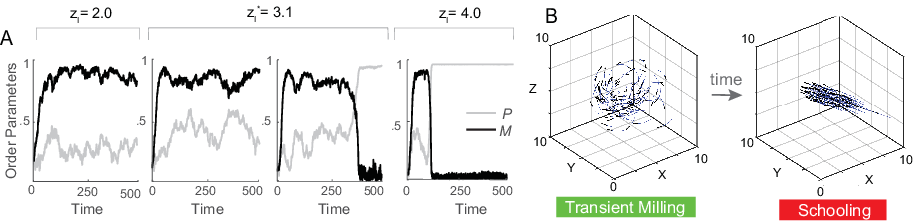}
    \caption{\footnotesize{\textbf{A.}
    Time evolution of $P$ and $M$ indicates stable milling at $z_l = 2$, bistable behavior at the transition point $z_l^\ast = 3.1$, where in one realization, the group remains in the milling state while in another realization, it transitions to schooling, and schooling at $z_l = 4$ after a short transience in the milling state. \textbf{B.} Snapshots showing that as time evolves, the group transitions from milling to schooling.  Parameter values $z_a = 10$, $\sigma = 0.01$.}}
    \label{fig:examples}
\end{figure}

\paragraph{Phase Space.} We systematically explored the parameter space $(z_l,z_a)$ by performing 50 Monte Carlo (MC) simulations for each combination $(z_l,z_a)$; each simulation lasted a total of 500 time units, and all simulations were conducted at noise level $\sigma = 0.01$ (Fig.~\ref{fig:phase diagram}E and F). At each point $(z_l,z_a)$, we averaged $P$ and $M$ over all 50 MC simulations.  To ensure that steady state was reached, we averaged over the last 50 time units of each simulation. Results are shown as a two-dimensional colormap over the $(z_l,z_a)$ space (Fig.~\ref{fig:phase diagram}E). 
Fragmentation arises at small values of $z_l$ and $z_a$, for which the school is unable to maintain group cohesion, resulting in disconnected clusters.  To distinguish this state, we calculated the average nearest-neighbor distance (NND) and corresponding standard deviation (Fig.~\ref{fig:phase diagram}F): an average NND that exceeds the range of interactions $z_a$ and exhibits high standard deviation is indicative of group fragmentation and dispersal. 
Swarming emerges at larger $z_a$, where individuals are attracted toward the group center without aligning with their neighbors. This state is characterized by low polarization $P$ and rotation $M$, indicating local aggregation but disorganized heading directions.  The swarming state transitions smoothly to milling as the radius of the alignment zone $z_l$ increases, but the milling state transitions abruptly to schooling at $z_l = z_l^\ast$, where $z_l^\ast\approx 3.1$ at $z_a = 10$. 
The smooth transition from swarming to milling is characterized by a smooth increase in the rotation order parameter $M$ while maintaining a low polarization $P$, as evident from the gradual change in color in Fig.~\ref{fig:phase diagram}E.  The sharp transition from milling to schooling involves an abrupt change in color from green (low $P$ and high $M$) to red (high $P$ and low $M$).  This transition is also accompanied by a decrease in the average NND (Fig.~\ref{fig:phase diagram}F, blue lines): schooling fish swim closer to each other.  Importantly, these transitions depend primarily on the alignment zone $z_l$ (Fig.~\ref{fig:phase diagram}E): beyond  $z_a \geq 7$, the attraction zone has little influence on the behavior of the group, indicating that the tendency to align with neighbors plays a dominant role in the collective phase transitions in cohesive groups.

\begin{figure}[!t]
    \centering
    \includegraphics[scale = 1]{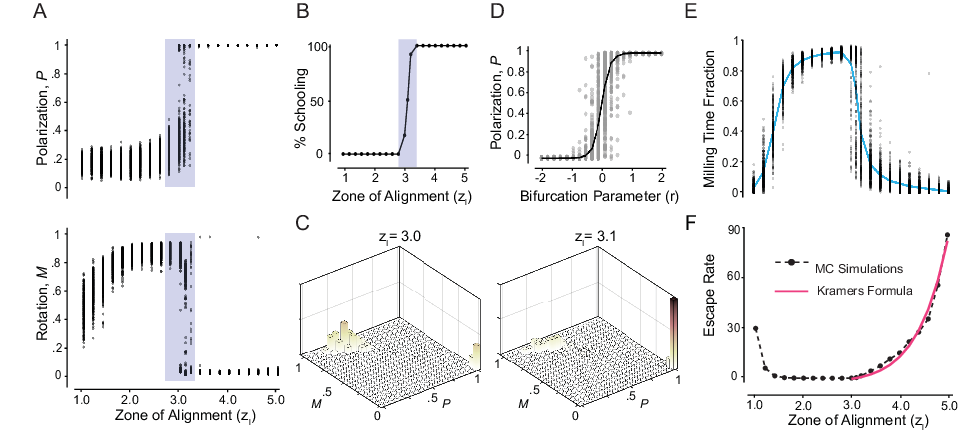}
    \caption{\footnotesize{\textbf{A.} $P$ and $M$ as a function of $z_l$: at each $z_l$, $P$ and $M$ values at steady state are averaged over 300 Monte Carlo simulations, with initial conditions randomly sampled in a three-dimensional unit cube. Behavior is bistable near the transition $z_l^\ast = 3.1$. 
    \textbf{B.} Percentage of MC simulations that reach schooling as a function of $z_l$. For $z_l \leq 2.8$, 100\% of the simulations converge to milling and for $z_l \geq 3.4$, 100\% converge to schooling. 
    \textbf{C.} Bimodal histograms of $P$ and $M$ at $z_l = 3$ and  $z_l^\ast = 3.1$ highlight that near the bifurcation, either milling or schooling are attainable. \textbf{E.} Fraction of time spent in the milling state as a function of $z_l$. 
    \textbf{D.} Transcritical bifurcation plot of the model in~\eqref{eq:normalform} showing 50 Monte Carlo simulations at $T = 10$ and $\sigma = 0.01$.
    \textbf{F.} Escape rate $\kappa$ from milling to schooling obtained numerically (black dashed line) and from a best fit of Kramers' formula in~\eqref{eq:kappa} (red line).  Parameter values $z_a = 10$, $\sigma = 0.01$, and $T = 500$. 
    }}
    \label{fig:bifurcation}
\end{figure}

\paragraph{Transition from Milling to Schooling.} 
We next took a closer look at the transition from milling to schooling. Given that $z_a$ has little effect on this transition, we fixed $z_a=10$ (Fig.~\ref{fig:phase diagram}E, dashed lines) for which the transition occurs at $z_l^\ast = 3.1$. 
To shed light on the dynamics across the transition point, we examined in Fig.~\ref{fig:examples}A the time evolution of $P$ and $M$:
at $z_l = 2$, the group maintains a consistently high value of $M$, reflecting its milling state. Interestingly, at the transition point $z_l^\ast = 3.1$, in one simulation the fish group converges to a persistent milling state for the remaining integration time while in another simulation, the group initially reaches the milling state, then transitions to the schooling state. 
Beyond the transition point, at $z_l = 4$, the group also experiences transient milling that gives way to schooling as time progresses. Snapshots highlighting the escape from milling to schooling as time progresses are shown in Fig.~\ref{fig:examples}B. The abrupt transition from milling to schooling at $z_l^\ast=3.1$ and the lingering in the milling state beyond this transition are hallmarks of a bifurcation -- a critical point where the milling state disappears or loses stability. 

To uncover the nature of this bifurcation, we systematically investigated the group dynamics by varying $z_l$ from 1.0 to 5.0 in increments of $\Delta z_l = 0.2$. For each value of $z_l$, we ran 300 Monte Carlo simulations for 500 time units each. Again, to ensure our analysis reflects the collective behavior at steady state, we averaged the values of $P$ and $M$ over the last 50 time units of each simulation. Fig.~\ref{fig:bifurcation}A shows  $P$ and $M$ as a function of $z_l$, emphasizing the existence of a transition from milling to schooling near $z_l^\ast = 3.1$. 
As $z_l$ approaches $z_l^\ast$ from the left, polarization $P$ sharply increases and rotational order $M$ sharply decreases, indicating a sudden decrease in the stability of the milling state.  Within a narrow range around $z_l^\ast$, the system appears to be bistable, with simulations converging to either milling or schooling. Beyond $z_l^\ast$, polarization $P$ becomes persistently high and rotational order $M$ persistently low, indicating a loss of stability of the milling state in favor of stable schooling. 
To better illustrate this transition, we plot in Fig.~\ref{fig:bifurcation}B the percentage of the MC simulations that converged to schooling as a function of $z_l$: bistable behavior, where some simulations converge to milling and others to schooling, is localized near the bifurcation value $z_l^\ast = 3.1$. 
Note that because the system is noisy, an exact bifurcation value is ambiguous. We thus considered the bifurcation point as the value $z_l^\ast = 3.1$, for which nearly half of the 300 MC simulations converged to schooling while the other half to milling. 
To underscore that bistability is a phenomenon localized near the bifurcation point, we computed histograms of $P$ and $M$ based on the 300 Monte Carlo simulations. For $z_l \leq 2.8$, all simulations converge to milling, leading to unimodal histograms centered at high $M$ and low $P$; For $z_l = 3.4$, they all converge to schooling, with unimodal histograms centered at high $P$ and low $M$ (histograms not shown for brevity). Near the transition, both milling and schooling co-exist, as evident from the bimodal histograms in Fig.~\ref{fig:bifurcation}C. Taken together, the results in Fig.~\ref{fig:bifurcation}A-C support the conclusion that the milling and schooling states exchange stability in a noisy transcritical bifurcation.

\paragraph{Signature of Noisy Transcritical Bifurcation.}   
A formal proof that the transition from milling to schooling at $z_l^\ast$ is due to a noisy transcritical bifurcation would require deriving the equations governing the time evolution of $P$ and $M$ and mapping them onto their ``normal form'' \cite{Man2020}.  Deriving these equations analytically from~\eqref{eq:vision} is challenging and approximating them numerically from simulation data is feasible but non-trivial~\cite{Kolpas2007,Huang2024}.
Instead of pursuing this approach, we sought a phenomenological model that describes the normal forms that $P$ and $M$ must obey to reproduce qualitatively the transition from milling to schooling reported in Fig.~\ref{fig:bifurcation}A,B. Since the behavior of $M$ is nearly a mirror opposite of that of $P$, it suffices to investigate the behavior of $P$, which we consider to follow the prototypical stochastic differential equation,
\begin{equation}
\begin{split}
dP & = -\nabla \mathcal{U} dt + \sigma dW_P = rP(1-P)dt + \sigma dW_P,
\label{eq:normalform}
\end{split}
\end{equation}
Here, we postulate an effective potential function of the form $\mathcal{U}  = -r( {P^2}/{2} - {P^3}/{3})$. The bifurcation parameter $r$ serves as proxy for the relative position from the bifurcation point $r = z_l - z_l^\ast$, $W_P$ is a standard Wiener process, with 
$\sigma > 0$ being the noise strength at the collective level, not necessarily equal to the noise strength at the individual fish level. In the deterministic version of this equation ($\sigma = 0$), $P$ admits two equilibria at $0$ and $1$ that persist for all $r$ values, but switch or exchange stability at the bifurcation point $r^\ast=0$: for $r<0$, $P=0$ (milling) is stable and $P=1$ (schooling) is unstable, and vice versa for $r>0$. This is the hallmark of a transcritical bifurcation (Fig.~\ref{fig:map}B).

At the bifurcation $r^\ast = 0$, the drift term vanishes and the effective potential $\mathcal{U} =0$ is flat: the dynamics are purely diffusive, with $P$ following a pure Brownian motion. Thus, depending on initial conditions, $P(t)$ converges to either $0$ or $1$. Away from the bifurcation, one can show formally that for $r>0$, the system is guaranteed to converge to $P=1$ (schooling) in the infinite time limit, and for $r<0$, convergence to $P=0$ (milling) is guaranteed.  Bistability is thus degenerate in the sense that, in infinite time, it arises only at the bifurcation point $r^\ast=0$.
In finite time simulations (Fig.~\ref{fig:bifurcation}D), noise enables initial conditions to `leak' near the bifurcation point from the pre-bifurcation stable equilibrium (milling) to the post-bifurcation state equilibrium (schooling).
This noise-induced bistability is qualitatively consistent with our numerical findings in Fig.~\ref{fig:bifurcation}; it is not to be confused with structural bistability~\cite{Huang2024}.

\paragraph{Residence Time and Escape Rate.} Next, we returned to our fish simulations with the aim of exploring the transient lingering in the milling state post the bifurcation point $z_l^\ast = 3.1$. We considered the time evolution of $P$ and $M$ of each of the 300 MC simulations in Fig.~\ref{fig:bifurcation}A. 
For each simulation, we computed the fraction of time $\tau=T_{\rm milling}/T$, where $T_{\rm milling}$ is the portion of time spent in the milling state. Plotting $\tau$ as a function of $z_l$ shows that as the group transitions smoothly from swarming to milling,  $\tau$ increases gradually to nearly 1, indicating that the group quickly converges to and remains in the milling state (Fig.~\ref{fig:bifurcation}E). However, as $z_l$ increases beyond the transition $z_l^\ast$ from milling to schooling, the fraction of time spent in the milling state decreases, until the transience in milling disappears entirely far beyond the bifurcation point. 
We next calculated the escape rate $\kappa$, which is inversely proportional to the average residence time $\langle \tau \rangle$ in the milling state (Fig.~\ref{fig:bifurcation}E, blue line).
As $z_l$ increases beyond $z_l^\ast$, the escape rate increases exponentially (Fig.~\ref{fig:bifurcation}F, dashed black line).  

To analyze this escape rate in the context of our phenomenological model in~\eqref{eq:normalform}, we employed Kramers' formula \cite{Kramers1924, Amir2023}, which describes the noise-induced escape rate from a potential well. 
According to Kramers' formula, the escape rate $\kappa$ from $P = 0$ to $P = 1$ as $r$ becomes positive is given by
\begin{equation}
\begin{split}
\kappa = \langle T_{\rm milling} \rangle^{-1} 
 = A \sqrt{|\mathcal{U}''|_{P=0}| \mathcal{U}''|_{P=1}} \exp\left( \dfrac{-2\Delta \mathcal{U}}{\sigma^2}  \right)
= A r \, e^{- r/3\sigma^2}.
\label{eq:kappa}
\end{split}
\end{equation}
Here, with $\mathcal{U}= -r( P^2/2 -P^3/3)$, we get that $\mathcal{U}'' = - r(1 -2P) = \mp r$ at $P=0$ and $1$, respectively, and the potential barrier $\Delta \mathcal{U} = \mathcal{U}|_{P=1} - \mathcal{U}|_{P=0} = {r}/{6}$. 
The proportionality constant $A$ and collective noise intensity $\sigma$ are obtained by fitting this formula to the numerical simulations in Fig.~\ref{fig:bifurcation}F.  With this numerical fit (Fig.~\ref{fig:bifurcation}F, red line), Kramers' formula captures the increase in the transition rate from milling to schooling past the bifurcation point.
These findings support our central argument that the bifurcation governing the transition from milling to schooling in Fig.~\ref{fig:bifurcation} is that of a noisy transcritical bifurcation of the form proposed in~\eqref{eq:normalform}.

\begin{figure}[!t]
    \centering
    \includegraphics[width=1\textwidth]{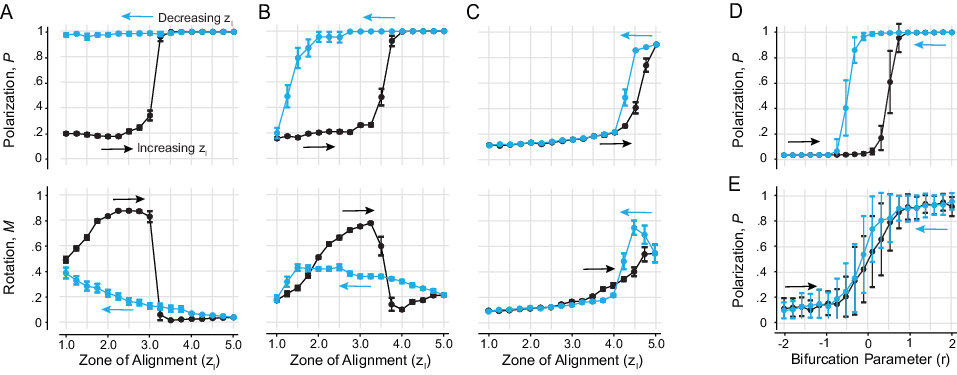}
    \caption{\footnotesize{Hysteresis plot at noise level \textbf{A.} $\sigma = 0.01$, \textbf{B.} $\sigma = 0.05$, \textbf{C.} $\sigma = 0.10$.  Each value is the average of 20 Monte Carlo simulations, $z_a = 10$, $\Delta z_l = 0.25$.  Each value of $z_l$ is run for 400 time units before it is increased (black line) or decreased (blue line). 
     \textbf{D.} Hysteresis plot based on the model in~\eqref{eq:normalform} at intermediate noise $\sigma = 0.01$ and \textbf{E.} high noise $\sigma = 0.20$; each value is the average of 50 Monte Carlo simulations and run for 10 time units before it is increased (black line) or decreased (blue line)
    }}  
    \label{fig:hysteresis}
\end{figure}

\paragraph{Noise-dependent Hysteresis.}
Thus far, our numerical analysis of group behavior relied on Monte Carlo simulations starting at random initial conditions. We next investigated scenarios where starting from an ordered state, we gradually shifted the individual's behavior, by increasing and decreasing the span of the alignment zone $z_l$. 
Specifically, starting from the steady state at $z_l=1.0$, we gradually increased $z_l$ from $1.0$ to $5.0$, using increments of $\Delta z_l= 0.25$ (Fig.~\ref{fig:hysteresis}, black markers). Then, starting from the steady state reached at $z_l=5.0$, we gradually decreased $z_l$ by the same increment (Fig.~\ref{fig:hysteresis}, blue markers). At each value of $z_l$, we performed 20 Monte Carlo simulations, each for a total duration of 400 time units.  We repeated this process at three distinct levels of noise $\sigma = 0.01$, $0.05$, and $0.10$, while keeping all other parameters the same. 

At low noise $\sigma = 0.01$ (Fig.~\ref{fig:hysteresis}A), the group transitions from swarming to milling to schooling as $z_l$ increases, with the transition from milling to schooling occurring near the bifurcation value $z_l^\ast = 3.1$ uncovered in Fig.~\ref{fig:bifurcation}. However, in the backward direction, the group remains locked in the schooling state and is unable to transition back to milling. These results suggest an asymmetric collective memory in one direction, that of decreasing $z_l$ from a schooling state, while the direction of increasing $z_l$ from a milling state seems to faithfully track the bifurcation diagram highlighted in Fig.~\ref{fig:bifurcation}A, without exhibiting strong memory effect.

At intermediate noise $\sigma = 0.05$ (Fig.~\ref{fig:hysteresis}B), we found a forward transition from swarming to milling to schooling.  However, compared to $\sigma = 0.01$, the transition occurs at a higher $z_l$ value.  In the backward direction, while the system is able to transition between states, the milling state is underrepresented and the system transitions directly from schooling to swarming. That is, at this noise level, the group exhibits a strong hysteresis, where the group patterns that form depend on the previous history of the group as reported in~\cite{Couzin2002,Couzin2009}. 

At even higher noise $\sigma = 0.10$ (Fig.~\ref{fig:hysteresis}C), the system exhibits no milling and transitions directly between swarming and schooling in both directions, albeit at a larger value of $z_l$, with negligible memory effects. That is, high noise levels overpower the stability of the milling state, removing it entirely and destroying the collective memory effect.

Taken together, these results show that the dependence of the system's collective behavior on prior history varies with noise level. The hysteresis effect is most pronounced at an intermediate noise value $\sigma = 0.05$, comparable to the one used in~\cite{Couzin2002,Couzin2009}.
Hysteresis here is induced by the noisy transcritical bifurcation uncovered in Fig.~\ref{fig:bifurcation} marking the transition from milling to schooling, where the nonlinear energy potential is of the form illustrated in Fig.~\ref{fig:map}B as opposed to the double well energy potential in Fig.~\ref{fig:map}A. To test this in the context of our model~\eqref{eq:normalform}, we systematically increased and decreased the bifurcation parameter $r$ in the same manner as in our simulations of the fish group, with each value held constant for 10 time units and averaged across 50 Monte Carlo simulations.  We repeated this process for two distinct collective noise levels $\sigma = 0.01$ and $0.20$. At $\sigma = 0.01$, we obtained a full hysteresis loop (Fig.~\ref{fig:hysteresis}D). At $\sigma = 0.20$, the hysteresis effect disappears, resulting in a direct transition between the two  states (Fig.~\ref{fig:hysteresis}E). This pattern mirrors the fish simulation results, corroborating that the collective memory effects, rather than arising from an intrinsically bistable system, correspond to a noisy transcritical bifurcation.

\section{Conclusions}

We revisited a popular model of schooling fish, referred to as the three-A model, where individuals are represented as self-propelled particles that interact via simple rules of avoidance, alignment, and attraction. Fixing the zone of avoidance to emulate the fish body length, we used Monte Carlo simulations to explore the two-dimensional parameter space defined by the alignment and attraction zones. We found that, beyond a range necessary for maintaining group cohesion, the zone of attraction has little effect on the collective transitions from swarming to milling then to schooling. These transitions are dictated by the alignment parameter.  Through a combination of numerical simulations and analytical tools from bifurcation theory, we showed that the transition from swarming to milling is smooth, but the transition from milling to schooling is abrupt and governed by a noisy transcritical bifurcation, where the two states, milling and schooling, intersect and exchange stability. Importantly, our findings show that this noisy transcritical bifurcation leads to hysteresis --  collective memory and dependence on past states -- thus clarifying a long-standing ambiguity about the origin of this hysteretic behavior in the three-A model~\cite{Couzin2002,Couzin2009}.

Beyond the three-A model, our work highlights that collective bistability could arise from distinct mechanisms and should not be treated as a single phenomenon.  Structural bistability of the type demonstrated in Fig.~\ref{fig:map}A arises when the collective dynamics is governed by a double-well effective energy potential. Structural bistability implies hysteresis, but the presence of hysteresis does not necessarily imply structural bistability. Indeed, in this work, we demonstrated that in the three-A model, hysteresis arises from a stochastic mechanism where noise delays the system’s response near the bifurcation from milling to schooling (Fig.~\ref{fig:map}B).

Distinguishing between the various mechanisms that can give rise to phase transitions and collective bistability is crucial for understanding and interpreting emergent behavior in animal groups. 
Consider for instance the intermittent, back-and-forth transitions between milling and polarized schooling observed empirically in fish schools under spatial confinement~\cite{Tunstrom2013,Lafoux2024} and varying light intensity~\cite{Lafoux2024}. Are these dynamic transitions governed by the same mechanisms?   Intermittency also arises in self-propelled particle models that account for rotational inertia of the individuals~\cite{Calovi2014,Wang2022} and for interactions with geometric boundaries~\cite{Huang2024}. To what extent do the mechanisms driving intermittency in these models reflect those at play in the biological fish schools?
To address these questions, a promising direction for future work is to  construct group-level effective potentials~\cite{Kolpas2007,Huang2024} from experimental data~\cite{Tunstrom2013,Lafoux2023,Lafoux2024} and compare these effective potentials to those that arise in self-propelled particle models~\cite{Couzin2002,Filella2018,Huang2024}. This would provide a quantitative framework for inferring classes of individual-level behavioral models that are consistent with empirical group-level dynamics.
 
More broadly, in addition to distinguishing between competing hypotheses about the individual control strategies and collective decision-making processes, mechanistic insights rooted in bifurcation theory allow us to predict the conditions under which collective transitions occur -- noise-induced transitions depend on the level of stochastic fluctuations in the group, whereas bifurcation-driven transitions require tuning of the individual behavior -- and to determine whether these transitions are reversible or path-dependent. 
Mechanistic understanding of these collective transitions also facilitates the transfer of insights across research domains, such as comparing emergent phenomena in animal groups to cognitive processes in the nervous system~\cite{Couzin2009}.

\paragraph{Acknowledgement.} Funding support provided by the NSF grants RAISE IOS-2034043 and CBET-210020, ONR grants N00014-22-1-2655 and N00014-19-1-2035, and NIH grant R01-HL153622 (all to E.K.).



\printbibliography

\end{document}